\documentclass{siamltex}

\newcommand{\figurecaption}[1]{\vspace{.2in}\caption{#1}}

\newcommand{\HH}[0]{{\cal H}}
\newcommand{\TT}[0]{{\cal T}}
\newcommand{\OT}[0]{\overline{{\cal T}}}

\newcommand{\AlgTexts}[0]{\cite{AHU74, AHU83, CLR91, HS76}}

\newcommand{\CiteAllStat}[0]{\cite{BCS83,
Cox75, Cox77, Cox78, Cox80, CS79, Sande77, Sande78, Sande84, Sande?}}

\newcounter{algcounterA}

\newtheorem{problem}{Problem}

\begin{document}

\title{Total Protection of Analytic Invariant Information 
\\ in Cross Tabulated Tables\thanks{A preliminary
version of this work appeared in Proc.~11th Annual Symposium on Theoretical
Aspects of Computer Science, Caen, France, February 24--26, 1994,
pp.~723--734.}} 

\author{Ming-Yang Kao\thanks{Department of Computer Science,
Duke University, Durham, NC 27708. Supported in part by NSF grants
MCS-8116678, DCR-8405478, and CCR-9101385.  Part of this work was done
while the author was at the Department of Computer Science, Yale
University, New Haven, Connecticut 06520.}}

\maketitle

\begin{abstract}
To protect sensitive information in a cross tabulated table, it
is a common practice to suppress some of the cells in the table.
An {\it analytic invariant} is a power series in terms of the
suppressed cells that has a unique feasible value and a
convergence radius equal to $+\infty$. Intuitively, the
information contained in an invariant is not protected even
though the values of the suppressed cells are not disclosed.
This paper gives an optimal linear-time algorithm for testing
whether there exist nontrivial analytic invariants in terms of
the suppressed cells in a given set of suppressed cells.  This
paper also presents NP-completeness results and an almost
linear-time algorithm for the problem of suppressing the minimum
number of cells in addition to the sensitive ones so that the
resulting table does not leak analytic invariant information
about a given set of suppressed cells.
\end{abstract}



\section{Introduction}\label{sec_intro}
Cross tabulated tables are used in a wide variety of documents to
organize and exhibit information, often with the values of some
cells suppressed in order to conceal sensitive information.
Concerned with the effectiveness of the practice of cell
suppression~\cite{Denning82}, statisticians have raised two
fundamental issues and developed computational heuristics to
various related problems {\CiteAllStat}.  The {\it detection}
issue is whether an adversary can deduce significant information
about the suppressed cells from the published data of a table.
The {\it protection} issue is how a table maker can suppress a
small number of cells in addition to the sensitive ones so that
the resulting table does not leak significant information.

This paper investigates the complexity of how to protect a broad class of
information contained in a two-dimensional table that publishes (1) the
values of all cells except a set of sensitive ones, which are {\it
suppressed}, and (2) an upper bound and a lower bound for each cell, and
(3) all row sums and column sums of the complete set of cells.  The cells
may have real or integer values. They may have different bounds, and the
bounds may be finite or infinite.  The upper bound of a cell should be
strictly greater than its lower bound; otherwise, the value of that cell is
immediately known even if that cell is suppressed.  The cells that are not
suppressed also have upper and lower bounds. These bounds are necessary
because some of the unsuppressed cells may later be suppressed to protect
the information in the sensitive cells.  (See
Figures~{\ref{figure_c_table_inv}} and {\ref{figure_p_table_inv}} for an
example of a complete table and its published version.)

An {\it unbounded feasible assignment} to a table is an assignment of
values to the suppressed cells such that each row or column adds up to its
published sum.  An {\it bounded} feasible assignment is an unbounded one
that also obeys the bounds of the suppressed cells.  An {\it analytic
function} of a table is a power series of the suppressed cells, each
regarded as a variable, such that the convergence radius is $\infty$
{\cite{Ahlfors79, Apostol74, Lang85, LS68, Royden88, Rudin75}}.  An {\it
analytic invariant} is an analytic function that has a unique value at all
the bounded feasible assignments.  If an analytic invariant is formed by a
linear combination of the suppressed cells, then it is called a {\it linear
invariant}~\cite{Kao.mli.acsc, KaoG93}.  Similarly, a suppressed cell is
called an {\it invariant cell}~\cite{Gusfield87, Gusfield88} if it is an
invariant by itself.  For instance, in the published table in
Figure~\ref{figure_p_table_inv}, let $X_{p,q}$ be the cell at row $p$ and
column $q$.  $X_{6,i}$ is an invariant because it is the only suppressed
cell in row $6$.  $X_{2,c}$ and $X_{3,c}$ are invariant cells because their
values are between 0 and 9.5, their sum is 19, and both cells are forced to
have the same unique value $9.5$. Consequently,
$(X_{3,c}{\cdot}X_{2,c}+0.5{\cdot}X_{2,c}-95)^2{\cdot}X_{1,b}
+\sin(X_{2,c}{\cdot}X_{2,a}-9.5{\cdot}X_{2,a})$ is also an invariant.

Intuitively, the information contained in an analytic invariant
is unprotected because its value can be uniquely deduced from the
published data.  In this paper, a set of suppressed cells is {\it
totally protected} if there exists no analytic invariant in terms
of the suppressed cells in the given set, except the trivial
invariant that contains no nonzero terms.  As the analytic power
series form a very broad family of mathematical functions, total
protection conceals from the adversary a very large class of
information.  This paper gives a very simple algorithm for
testing whether a given set of suppress cells is totally
protected.  When a graph representation, called the {\it
suppressed graph}, of a table is given as input, this algorithm
runs in optimal $O(m+n)$ time, where $m$ is the number of
suppressed cells and $n$ is the total number of rows and columns.
This paper also considers the problem of computing and
suppressing the minimum number of additional cells so that a
given set of original suppressed cells becomes totally protected.
This problem is shown to be NP-complete.  For a large class of
tables, this optimal suppression problem can be solved in
$O((m+n){\cdot}\alpha(n,m+n))$ time, where $\alpha$ is an
Ackerman's inverse function and its value is practically a small
constant~\AlgTexts.  Moreover, for this class of tables, every
optimal set of cells for additional suppression forms a spanning
forest of some sort. As a consequence, at most $n-1$ additional
cells need to be suppressed to achieve the total protection of a
given set of original suppressed cells.  As the size of a table
may grow quadratically in $n$, the suppression of $n-1$
additional cells is a negligible price to pay for total
protection for a reasonably large table.

Previously, four other levels of data security have been considered that
protect information contained, respectively, in individual suppressed
cells~\cite{Gusfield87, Gusfield88}, in a row or column as a whole, in a
set of $k$ rows or $k$ columns as a whole, and in a table as a
whole~\cite{Kao95.protection}.  These four levels of data security and
total protection differ in two major aspects.  First, these four levels of
data security primarily protect information expressible as linear
invariants, whereas total protection protects the much broader class of
analytic invariant information. Second, these four levels of data security
emphasize protecting regular regions of a table, whereas total protection
protects any given set of suppressed cells and is more flexible.  These
four levels of data security and total protection share some interesting
similarities.  As total protection corresponds to spanning forests in
suppressed graphs, these four levels of data security are equivalent to
some forms of 2-edge connectivity~\cite{Gusfield87, Gusfield88}, 2-vertex
connectivity, $k$-vertex connectivity and graph
completeness~\cite{Kao95.protection}.  In this paper, the NP-completeness
results and efficient algorithms for total protection rely heavily on its
graph characterizations.  Similarly, the equivalence characterizations of
these four levels of data security have been key in obtaining efficient
algorithms {\cite{Gusfield87, Gusfield88, Kao95.protection}} and
NP-completeness proofs {\cite{Kao95.protection}} for various detection and
protection problems.

\begin{figure}[p]
\begin{center}
      \newcommand{\tabone  }[1]{\raisebox{0.18cm}{\rule{0cm}{.55cm}#1}}
      \newcommand{\tabtwo  }[1]{\raisebox{0.18cm}{\rule{0cm}{.55cm}#1}}
      \newcommand{\maxone  }[0]{\tabone{9.5}}
      \newcommand{\midone  }[0]{\tabone{4.5}}
      \newcommand{\minone  }[0]{\tabone{0}}
      \newcommand{\opeone  }[0]{\tabone{1.5}}
      \newcommand{\opetwo  }[0]{\tabone{2}}
      \newcommand{\opethr  }[0]{\tabone{3}}
      \newcommand{\opefou  }[0]{\tabone{4}}
      \newcommand{\opefiv  }[0]{\tabone{5.5}}
      \newcommand{\opesix  }[0]{\tabone{6}}
      \newcommand{\opesev  }[0]{\tabone{7}}
      \newcommand{\opeeig  }[0]{\tabone{8}}
      \newcommand{\rowtop  }[2]{#1         & a       & b       & c       & d       & e       & f       & g       & h       & i       & #2         }
      \newcommand{\rowone  }[0]{\tabtwo{1} & \maxone & \midone & \opeone & \opesev & \opeone & \opeone & \opefiv & \opetwo & \opethr & \tabtwo{36.0}}
      \newcommand{\rowtwo  }[0]{\tabtwo{2} & \midone & \maxone & \maxone & \midone & \midone & \maxone & \maxone & \maxone & \midone & \tabtwo{65.5}}
      \newcommand{\rowthree}[0]{\tabtwo{3} & \opesix & \opeone & \maxone & \minone & \maxone & \opesix & \opefiv & \opetwo & \opefiv & \tabtwo{45.5}}
      \newcommand{\rowfour }[0]{\tabtwo{4} & \opetwo & \opeone & \opefou & \opesev & \opeone & \midone & \maxone & \opefiv & \opetwo & \tabtwo{37.5}}
      \newcommand{\rowfive }[0]{\tabtwo{5} & \opeone & \opefiv & \opefou & \opesix & \opefiv & \minone & \minone & \midone & \maxone & \tabtwo{36.5}}
      \newcommand{\rowsix  }[0]{\tabtwo{6} & \opetwo & \opethr & \opethr & \opefou & \opesix & \opefiv & \opetwo & \opetwo & \maxone & \tabtwo{37.0}}
      \newcommand{\rowbot  }[1]{#1        & 25.5   & 25.5     & 31.5    & 28.5
  & 28.5    & 27.0    & 32.0    & 25.5    & 34.0   &            }
      \newcommand{\RowColumnIndex}[0]{\parbox{1.0cm }{\rule[-.10cm]{0cm}{1.00cm}\shortstack{row\\column\\index}}}
      \newcommand{\RowSum        }[0]{\hspace{.12cm }\parbox{0.57cm}{\shortstack{row\\sum}}}
      \newcommand{\ColumnSum     }[0]{\parbox{1.01cm}{\rule[-.18cm]{0cm}{0.75cm}\shortstack{column\\sum       }}}

      \footnotesize
\begin{tabular}{|c||c|c|c|c|c|c|c|c|c||c|}                 
            \hline \rowtop{\RowColumnIndex}{\RowSum} \\
            \hline \hline \rowone                    \\
            \hline \rowtwo                           \\
            \hline \rowthree                         \\
            \hline \rowfour                          \\
            \hline \rowfive                          \\
            \hline \rowsix                           \\
            \hline \hline \rowbot{\ColumnSum}        \\
            \hline
\end{tabular}
\end{center}
\figurecaption{A Complete Table.}
\label{figure_c_table_inv}
\end{figure}

\begin{figure}[p]
\begin{center}
      \newcommand{\tabone  }[1]{\raisebox{0.18cm}{\rule{0cm}{.55cm}#1}}
      \newcommand{\tabtwo  }[1]{\raisebox{0.18cm}{\rule{0cm}{.55cm}#1}}
      \newcommand{\maxone  }[0]{ }
      \newcommand{\midone  }[0]{ } 
      \newcommand{\minone  }[0]{ } 
      \newcommand{\opeone  }[0]{\tabone{1.5}}
      \newcommand{\opetwo  }[0]{\tabone{2}}
      \newcommand{\opethr  }[0]{\tabone{3}}
      \newcommand{\opefou  }[0]{\tabone{4}}
      \newcommand{\opefiv  }[0]{\tabone{5.5}}
      \newcommand{\opesix  }[0]{\tabone{6}}
      \newcommand{\opesev  }[0]{\tabone{7}}
      \newcommand{\opeeig  }[0]{\tabone{8}}
      \newcommand{\rowtop  }[2]{#1         & a       & b       & c       & d       & e       & f       & g       & h       & i       & #2         }
      \newcommand{\rowone  }[0]{\tabtwo{1} & \maxone & \midone & \opeone & \opesev & \opeone & \opeone & \opefiv & \opetwo & \opethr & \tabtwo{36.0}}
      \newcommand{\rowtwo  }[0]{\tabtwo{2} & \midone & \maxone & \maxone & \midone & \midone & \maxone & \maxone & \maxone & \midone & \tabtwo{65.5}}
      \newcommand{\rowthree}[0]{\tabtwo{3} & \opesix & \opeone & \maxone & \minone & \maxone & \opesix & \opefiv & \opetwo & \opefiv & \tabtwo{45.5}}
      \newcommand{\rowfour }[0]{\tabtwo{4} & \opetwo & \opeone & \opefou & \opesev & \opeone & \midone & \maxone & \opefiv & \opetwo & \tabtwo{37.5}}
      \newcommand{\rowfive }[0]{\tabtwo{5} & \opeone & \opefiv & \opefou & \opesix & \opefiv & \minone & \minone & \midone & \maxone & \tabtwo{36.5}}
      \newcommand{\rowsix  }[0]{\tabtwo{6} & \opetwo & \opethr & \opethr & \opefou & \opesix & \opefiv & \opetwo & \opetwo & \maxone & \tabtwo{37.0}}
      \newcommand{\rowbot  }[1]{#1        & 25.5   & 25.5     & 31.5    & 28.5
  & 28.5    & 27.0    & 32.0    & 25.5    & 34.0    &            }
      \newcommand{\RowColumnIndex}[0]{\parbox{1.0cm }{\rule[-.10cm]{0cm}{1.00cm}\shortstack{row\\column\\index}}}
      \newcommand{\RowSum        }[0]{\hspace{.12cm }\parbox{0.57cm}{\shortstack{row\\sum}}}
      \newcommand{\ColumnSum     }[0]{\parbox{1.01cm}{\rule[-.18cm]{0cm}{0.75cm}\shortstack{column\\sum       }}}

      \footnotesize
\begin{tabular}{|c||c|c|c|c|c|c|c|c|c||c|}                 
            \hline \rowtop{\RowColumnIndex}{\RowSum} \\
            \hline \hline \rowone                    \\
            \hline \rowtwo                           \\
            \hline \rowthree                         \\
            \hline \rowfour                          \\
            \hline \rowfive                          \\
            \hline \rowsix                           \\
            \hline \hline \rowbot{\ColumnSum}        \\
            \hline
\end{tabular}

\parbox{4in}{\vspace{.2in} \small 
Note: Let $X_{p,q}$ denote the cell at row $p$ and column $q$.
The lower and upper bounds for all suppressed cells except
$X_{2,c}$ and $X_{3,c}$ are $-\infty$ and $+\infty$.  The lower
and upper bounds for $X_{2,c}$ and $X_{3,c}$ are 0 and 9.5.}
\end{center}
\figurecaption{A Published Table.}
\label{figure_p_table_inv}
\end{figure}

Section~\ref{sec_basic} discusses basic concepts.  Section~\ref{sec_total}
formally defines the notion of total protection and gives a linear-time
algorithm to test for this notion. Sections~\ref{sec_npc} and
{\ref{sec_linear}} give NP-completeness results and efficient algorithms
for optimal suppression problems of total protection.
Section~\ref{sec_extension} concludes this paper with discussions.

\section{Basics of two-dimensional tables}\label{sec_basic}
This section discusses basic relationships between tables and graphs.

A {\it mixed} graph is one that may contain both undirected and
directed edges.  A {\it traversable} cycle or path in a mixed
graph is one that can be traversed along the directions of its
edges.  A {\it direction-blind} cycle or path is one that can be
traversed if the directions of its edges are disregarded.  The
word direction-blind is often omitted for brevity.
A mixed graph is {\it connected} (respectively, {\it strongly
connected}) if each pair of vertices are contained in a
direction-blind path (respectively, traversable cycle).  A {\it
connected component} (respectively, {\it strongly connected
component}) of a mixed graph is a maximal subgraph that is
connected (respectively, strongly connected).
A set of edges in a mixed graph is an {\it edge cut} if its
removal disconnects one or more connected components of that
graph.  An edge cut is a {\it minimal} one if it has no proper
subset that is also an edge cut.

\begin{figure}[p]
\newcommand{\tabone  }[1]{\raisebox{0.18cm}{\rule{0cm}{.55cm}#1}}
\newcommand{\tabtwo  }[1]{\raisebox{0.18cm}{\rule{0cm}{.55cm}#1}}
\newcommand{\maxone  }[0]{\tabone{\framebox[0.5cm]{\footnotesize 9}}}
\newcommand{\midone  }[0]{\tabone{\framebox[0.5cm]{\footnotesize 5}}}
\newcommand{\minone  }[0]{\tabone{\framebox[0.5cm]{\footnotesize 0}}}
\newcommand{\opeone  }[0]{\tabone{1}}
\newcommand{\opesix  }[0]{\tabone{6}}
\newcommand{\rowtop  }[2]{#1        &a      &b      &c      &#2         }
\newcommand{\rowone  }[0]{\tabtwo{1}&\minone&\maxone&\opeone&\tabtwo{10}}
\newcommand{\rowtwo  }[0]{\tabtwo{2}&\maxone&\maxone&\minone&\tabtwo{18}}
\newcommand{\rowthree}[0]{\tabtwo{3}&\opesix&\minone&\midone&\tabtwo{11}}
\newcommand{\rowbot  }[1]{#1        &15     &18     &6      &           }
\newcommand{\RowColumnIndex}[0]{\parbox{1.0cm }{\rule[-.10cm]{0cm}{1.00cm}\shortstack{row\\column\\index}}}
\newcommand{\RowSum        }[0]{\hspace{.12cm }\parbox{0.57cm}{\shortstack{row\\sum}}}
\newcommand{\ColumnSum     }[0]{\parbox{1.01cm}{\rule[-.18cm]{0cm}{0.75cm}\shortstack{column\\sum       }}}

            \begin{center}                   
                  \footnotesize
                  \begin{tabular}{|c||c|c|c||c|}                 
                        \hline \rowtop{\RowColumnIndex}{\RowSum} \\
                        \hline \hline \rowone                    \\
                        \hline \rowtwo                           \\
                        \hline \rowthree                         \\
                        \hline \hline \rowbot{\ColumnSum}        \\
                        \hline
                  \end{tabular}
            \end{center}

            \vspace{0.6in}
            \begin{center}
                  \begin{picture}(160,60)(0,0)
                         \setlength{\unitlength}{.1cm}
                         \multiput(12,12)(16,0){3}{\circle*{4}}  
                         \multiput(12,28)(16,0){3}{\circle*{4}}  
                         \put(12,12){\line  ( 1, 0){32}}
                         \put(12,12){\vector(+1, 0){14}}
                         \put(28,12){\vector(+1, 0){14}}
                         \put(12,28){\line  ( 1, 0){32}}
                         \put(44,28){\vector(-1, 0){14}}
                         \put(28,28){\vector(-1, 0){14}}
                         \put(12,12){\line  ( 0, 1){16}}
                         \put(12,28){\vector( 0,-1){14}}
                         \put(28,12){\line  ( 0, 1){16}}
                         \put(28,28){\vector( 0,-1){14}}
                         \put(44,12){\line  ( 0, 1){16}}
                         \put(10, 6){$C_a$}
                         \put(26, 6){$R_2$}
                         \put(42, 6){$C_c$}
                         \put(10,31){$R_1$}
                         \put(26,31){$C_b$}
                         \put(42,31){$R_3$}
                   \end{picture}
            \end{center}

\begin{center}
\parbox{5in}{\small      
In the above $3 \times 3$ table, the number in each cell is the value of that
cell.  A cell with a box is a suppressed cell.  The lower and upper bounds of
the suppressed cells are 0 and 9.  The graph below the table is the suppressed
graph of the table.  Vertex $R_p$ corresponds to row $p$, and vertex $C_q$ to
column $q$.}
\end{center}
\figurecaption{A Table and Its Suppressed Graph.}
\label{figure_suppressed_graph}
\end{figure}

From this point onwards, let $\TT$ be a table, and let
$\HH'=(A,B,E')$ and $\HH=(A,B,E)$ be the bipartite mixed graphs
constructed below.  $\HH'$ and $\HH$ are called the {\it total
graph} and the {\it suppressed graph} of $\TT$,
respectively~\cite{Gusfield88}.
For each row (respectively, column) of $\TT$, there is a unique
vertex in $A$ (respectively, $B$). This vertex is called a {\it
row} (respectively, {\it column}) vertex.
For each cell $X_{i,j}$ at row $i$ and column $j$ in $\TT$, there
is a unique edge $e$ in $E$ between the vertices of row $i$ and
column $j$.
If the value of $X_{i,j}$ is strictly between its bounds, then
$e$ is undirected. Otherwise, if the value is equal to the lower
(respectively, upper) bound, then $e$ is directed towards to its
column (respectively, row) endpoint.
Note that $\HH'$ is a {\it complete} bipartite mixed graph, i.e.,
there is exactly one edge between each pair of vertices from the
two vertex sets of the graph.
The graph $\HH$ is the subgraph of $\HH'$ whose edge set consists
of only those corresponding to the suppressed cells of $\TT$.
 Figure~\ref{figure_suppressed_graph} illustrates a table and its
suppressed graph.  
For convenience, a row or column of $\TT$ will be regarded as a
vertex in $\HH$ and a cell as an edge, and vice versa.

\begin{theorem}[\cite{Gusfield88}]\label{thm_invariant_cell} 
A suppressed cell of $\TT$ is an invariant cell if and only if it
is not in an edge-simple traversable cycle of $\HH$.
\end{theorem}

The {\it effective area} of an analytic function $F$ of $\TT$,
denoted by $EA(F)$, is the set of variables in the nonzero terms
of $F$.  The function $F$ is called {\it nonzero} if
$EA(F)\neq\emptyset$. Note that because the convergence radius of
$F$ is $\infty$, $EA(F)$ is independent of the point at which $F$
is expanded into a power series.

\begin{theorem}[\cite{Kao.mli.acsc}]\label{thm_mli}
For every minimal edge cut $Y$ of a strongly connected component
of $\HH$, $\TT$ has a linear invariant $F$ with $EA(F)$ = $Y$.
\end{theorem}

The {\it bounded kernel} (respectively, {\it unbounded kernel})
of $\TT$, denoted by $BK(\TT)$ (respectively, $UK(\TT)$), is the
real vector space consisting of all linear combinations of $x-y$,
where $x$ and $y$ are arbitrary bounded (respectively, unbounded)
feasible assignments of $\TT$.

Because $\HH$ is bipartite, every cycle of $\HH$ is of even length. Thus,
the edges of an edge-simple direction-blind cycle of $\HH$ can be
alternately labeled with $+1$ and $-1$. Such a labeling is called a {\it
direction-blind labeling}.  A direction-blindly labeled cycle is regarded
as an assignment to the suppressed cells of $\TT$.  If the corresponding
edge of a suppressed cell is in the given cycle, then the value assigned to
that cell is the label of that edge; otherwise, the value is 0.  Note that
this assignment needs not be an unbounded feasible assignment of $\TT$.

\begin{theorem}[\cite{KaoG93}]\label{thm_uk_bk_lem_unbounded_embedding}
\begin{enumerate}
\item
$UK(\TT) = BK(\TT)$ if every connected component of $\HH$ is
strongly connected.
\item
Every direction-blindly labeled cycle of $\HH$ is a vector in
$UK(\TT)$.
\end{enumerate}
\end{theorem}

\section{Total protection}\label{sec_total}
A set $Q$ of suppressed cells of $\TT$ is {\it totally protected}
in $\TT$ if there is no nonzero analytic invariant $F$ of $\TT$
with $EA(F) \subseteq Q$.
The goal of total protection can be better understood by
considering $Q$ as the set of suppressed cells that contain
sensitive data.  The total protection of $Q$ means that no
precise analytic information about these data, not even their row
and column sums, can be deduced from the published data of $\TT$.
As analytic power series form a very large class of functions in
mathematical sciences, this notion of protection requires a large
class of information about $Q$ to be concealed from the
adversary.

The next lemma and theorem characterize the notion of total
protection in graph concepts.

\begin{lemma}\label{lem_analytic_cut}
If $F$ is a nonzero analytic invariant of $\TT$ such that the
edges in $EA(F)$ are contained in the strongly connected
components of $\HH$, then for some strongly connected component
$D$ of $\HH$, $EA(F) \cap D$ is an edge cut of $D$.
\end{lemma}

{\it Remark.} The converse of this lemma is not true; for a counter
example, consider the linear combination $X_{1,a}+2{\cdot}X_{1,b}$ for the
table in Figure~\ref{figure_suppressed_graph}.  Also, if $F$ is a nonzero
linear invariant, then for every strongly connected component $D$ of $\HH$,
the set $D \cap EA(F)$ is either empty or is an edge cut of
$D$~\cite{Kao.mli.acsc}.

\begin{proof}
Let $\TT_s$ be the table constructed from $\TT$ by also publishing
the suppressed cells that are not in the strongly connected
components of $\HH$.  By Theorem~\ref{thm_invariant_cell}, $F$
remains a nonzero analytic function of $\TT_s$.  Also, the
connected components of the suppressed graph $\HH_s$ of $\TT_s$ are
the strongly connected components of $\HH$.  Thus, to prove the
lemma, it suffices to prove it for $\TT_s$, $\HH_s$, and $F$.

Let $x_0$ be a fixed bounded feasible assignment of $\TT_s$.  Let
$K=\{x-x_0| x$ is a bounded feasible assignment of $\TT_s\}$.  Since $F$ is
an analytic invariant of $\TT_s$, the function $G(x) = F(x)-F(x_0)$ is an
analytic invariant of $\TT_s$ with $EA(G)=EA(F)$ and its value is zero over
$x_0+K$.  Because $K$ contains a nonempty open subset of $BK(\TT_s)$, $G$
is zero over $x_0+BK(\TT_s)$.  By
Theorem~\ref{thm_uk_bk_lem_unbounded_embedding}(1) and the strong
connectivity of the connected components of $\HH_s$, $BK(\TT_s)=UK(\TT_s)$
and $G$ is zero over $x_0+UK(\TT_s)$.  Thus, it suffices to show that if
$D-EA(F)$ is connected for all connected components $D$ of $\HH_s$, then
$G(x_0+z_0) \not= 0$ for some $z_0
\in UK(\TT_s)$.  To construct $z_0$, let
$EA(G)=\{e_1,\ldots,e_k\}$.  Let $D_i$ be the connected component
of $\HH_s$ that contains $e_i$. By the connectivity of $D_i-EA(F)$,
there is a vertex-simple path $P_i$ in $D_i-EA(F)$ between the
endpoints of $e_i$.  Let $C_i$ be the vertex-simple cycle formed
by $e_i$ and $P_i$. Next, direction-blindly label $C_i$ with
$e_i$ labeled $+1$.  Since $G$ is a nonzero power series,
$G(x_0+y_0) \not = 0$ for some vector $y_0$. Note that $y_0$ is
not necessarily in $UK(\TT_s)$.  So, let $z_0 = \sum_{i=1}^k
h_i{\cdot}C_i$, where $h_i$ is the component of $y_0$ at variable
$e_i$.  Then, by Theorem~\ref{thm_uk_bk_lem_unbounded_embedding}(2), $z_0 \in
UK(\TT_s)$.  Because $P_i$ is in $\HH_s-EA(F)$, $e_i$ appears only in
the term $C_i$ in \(\sum_{i=1}^k h_i{\cdot}C_i\).  Thus $z_0$ and
$y_0$ have the same component values at the variables in $EA(G)$.
Since the variables not in $EA(G)$ do not appear in any expansion
of $G$, $G(x_0+z_0)=G(x_0+y_0) \neq 0$, proving the lemma.
\end{proof}

\begin{theorem}\label{thm_test}
A set $Q$ of suppressed cells is totally protected in $\TT$ if
and only if the two statements below are both true:
\begin{enumerate}
\item 
The edges in $Q$ are contained in the strongly connected
components of $\HH$.
\item 
For each strongly connected component $D$ of $\HH$, the graph
$D-Q$ is connected.
\end{enumerate}
\end{theorem}
\begin{proof} 
It is equivalent to show that $Q$ is not totally protected if and
only if $Q$ contains some edges not in the strongly connected
components of $\HH$ or for some strongly connected component $D$
of $\HH$, the graph $D-Q$ is not connected.  The $\Rightarrow$
direction follows from Lemma~\ref{lem_analytic_cut}.  As for the
$\Leftarrow$ direction, if $Q$ contains some edges not in the
strongly connected components of $\HH$, then by
Theorem~\ref{thm_invariant_cell}, $Q$ contains some invariant
cells of $\TT$ and thus cannot be totally protected. If for some
strongly connected component $D$ of $\HH$, the graph $D-Q$ is not
connected, then some subset $Y$ of $Q$ is a minimal edge cut of
$D$.  By Theorem~\ref{thm_mli}, $\TT$ has a linear invariant $F$
with $EA(F) = Y$ and thus $Q$ is not totally protected.
\end{proof}

This paper investigates the following two problems concerning how
to achieve total protection.
\begin{problem}[Protection Test]\label{problem_test}\rm
\begin{itemize}
\item Input:
The suppressed graph $\HH$ and a set $Q$ of suppressed cells of a
table $\TT$.
\item Output:
Is $Q$ totally protected in $\TT$?
\end{itemize}
\end{problem}
\begin{theorem}
Problem~\ref{problem_test} can be solved in linear time in the
size of $\HH$.
\end{theorem}
\begin{proof}
This problem can be solved within the desired time bound by means
of Theorem~\ref{thm_test} and linear-time algorithms for
computing connected components and strongly connected components
\AlgTexts.
\end{proof}

\begin{problem}[Optimal Suppression]\label{problem_table_general}\rm
\begin{itemize}
\item Input:
A table $\TT$, a subset $Q$ of $E$, and an integer $p \geq 0$, where $E$ is
the set of all suppressed cells in $\TT$.
\item Output:
Is there a set $P$ consisting of at most $p$ published cells of $\TT$ such
that $Q$ is totally protected in the table $\OT$ formed by $\TT$ with the
cells in $P$ also suppressed?
\end{itemize}
\end{problem}

This problem is clearly in NP.  Section~\ref{sec_npc} shows that
this problem with $Q=E$ is NP-complete.  In contrast,
Section~\ref{sec_linear} proves that if the total graph of $\TT$
is undirected, then this problem with general $Q$ can be solved
in almost linear time.

\section{NP-completeness of optimal suppression}\label{sec_npc}
Throughout this section, the total graph of $\TT$ may or may not
be undirected.
    
\begin{theorem}\label{thm_npc}
Problem~\ref{problem_table_general} with $Q=E$ is NP-complete.
\end{theorem}

To prove this theorem, the idea is to first transform
Problem~\ref{problem_table_general} with $Q=E$ to the following
graph problem and then prove the NP-completeness of the graph
problem.

\begin{problem}\label{problem_graph_general}\rm
\begin{itemize}
\item Input:
A complete bipartite mixed graph $\HH'=(A,B,E')$, a subgraph
$\HH=(A,B,E)$, and an integer $p\geq0$.
\item Output: 
Does any set $P$ of at most $p$ edges in $E'-E$ hold the following two
properties?  
\end{itemize}
\begin{enumerate}
\item[] Property N1: 
Every connected component of $(A, B, E \cup P)$ is strongly connected.
\item[] Property N2:
The vertices of each connected component of $\HH$ are connected in
$(A,B,P)$, i.e., contained in a connected component in $(A,B,P)$.
\end{enumerate}
\end{problem}

\begin{lemma}\label{lemma_equiv_general}
Problem \ref{problem_table_general} with $Q=E$ and Problem
{\ref{problem_graph_general}} can be reduced to each other in
linear time.
\end{lemma}
\begin{proof} 
Given an instance $\TT$ and $p$ of
Problem~\ref{problem_table_general} with $Q=E$, the desired
instance of Problem~\ref{problem_graph_general} is the total
graph $\HH'=(A,B,E')$ and the suppressed graph $\HH=(A,B,E)$ of
$\TT$, and $p$ itself.  This transformation can easily be
computed in linear time.  There are two directions to show that
it reduces Problem~\ref{problem_table_general} to
Problem~\ref{problem_graph_general}. Assume that $P$ is a desired
set for Problem~\ref{problem_graph_general}.  By 
 Property N1, Statement 1 in Theorem~\ref{thm_test} is
true.  Also, every strongly connected component of $(A,B,E
\cup P)$ is a union of edge-disjoint connected components in
$\HH$ and $(A,B,P)$.  Therefore, by  Property N2,
Statement 2 of Theorem~\ref{thm_test} holds. As a result, $P$
itself is a desired set for Problem~\ref{problem_table_general}.
On the other hand, assume that $P$ is a desired set for
Problem~\ref{problem_table_general}.  Let $P'$ be the set of all
edges in $P$ that are also in the strongly connected components
of $(A,B,E \cup P)$.  By Statement 1 of Theorem~\ref{thm_test}
and the total protection of $E$ in $\OT$, the connected
components of $(A,B,E \cup P')$ are the strongly connected
components of $(A,B,E \cup P)$. Thus, $P'$ holds 
 Property N1.  Next, because a connected component of
$\HH$ is included in a strongly connected component of $(A,B,E
\cup P')$, by Statement 2 of Theorem~\ref{thm_test}, $P'$ also
holds  Property N2 and thus is a desired set for
Problem~\ref{problem_graph_general}.

Given an instance $\HH'$, $\HH$, and $p$ of
Problem~\ref{problem_graph_general}, the desired instance of
Problem~\ref{problem_table_general} with $Q=E$ is $p$ itself and
the table defined as follows.  For each vertex in $A$
(respectively, $B$), there is a row (respectively, column).  The
upper and lower bounds for each cell are 2 and 0. For each edge
$e$ in $E'$, its corresponding cell is at the row and column
corresponding to its endpoints.  The value of that cell is 1
(respectively, 0 and 2) if $e$ is undirected (respectively,
directed from ${A}$ to ${B}$, or directed from ${B}$ to ${A}$).
For each edge $e$ in $\HH$, its corresponding cell is suppressed.
Note that the total and suppressed graphs of this table are
$\HH'$ and $\HH$ themselves.  Thus, the remaining proof details
for this reduction are essentially the same as for the other
reduction.
\end{proof}

Both Problem~\ref{problem_table_general} with $Q=E$ and
Problem~\ref{problem_graph_general} are clearly in NP. To prove
their completeness in NP, by Lemma~\ref{lemma_equiv_general} it
suffices to reduce the following NP-complete problem to
Problem~\ref{problem_graph_general}.

\begin{problem}[Hitting Set~\cite{GJ79}]\label{problem_hit}
\rm
\begin{itemize}
\item Input:
A finite set $S$, a nonempty family $W$ of subsets of $S$, and an
integer $h \geq 0$.
\item Output:
Is there a subset $S'$ of $S$ such that $|S'| \leq h$ and
$S'$ contains at least one element in each set in $W$?
\end{itemize}
\end{problem}

Given an instance $S=\{s_1,\ldots,s_q\}$, $W=\{S_1,\ldots,S_r\}$,
$h$ of Problem~\ref{problem_hit}, an instance
$\HH'=(A,B,E'),\HH=(A,B,E),p$ of
Problem~\ref{problem_graph_general} is constructed as follows:
\begin{itemize}
\item Rule 1:
Let $A=\{a_0,a_1,\ldots,a_q\}$. The vertices
$a_1,\ldots,a_q$ correspond to $s_1,\ldots,s_q$, but
$a_0$ corresponds to no $s_i$.

\item Rule 2:
Let $B=\{b_0,b_1,\ldots,b_r\}$.  The vertices $b_1,\ldots,b_r$
correspond to $S_1,\ldots,S_r$ of $S$, but $b_0$ corresponds to
no $S_j$.

\item Rule 3:
Let $E'$ be the union of the following sets of edges:
\begin{enumerate}
\item 
$\{{b_0}\rightarrow{a_0}\}$.

\item 
$\{{a_0}\rightarrow{b_j} \mid \forall$ $j$ with $1 \leq j \leq
r\}$.

\item 
$\{{a_i}\rightarrow{b_0} \mid \forall$ $i$ with $1 \leq i \leq
q\}$.

\item 
$\{{b_j}\rightarrow{a_i} \mid \forall$ $s_i$ and $S_j$ with $s_i \in
S_j\}$.

\item
$\{{a_i}\rightarrow{b_j} \mid \forall$ $s_i$ and $S_j$ with $s_i
\not\in S_j\}$.

\end{enumerate}

\item Rule 4:
Let
$E=\{{a_0}\rightarrow{b_1},\ldots,{a_0}\rightarrow{b_r}\}$.
 
\item  Rule 5:
Let $p=h+r+1$.
\end{itemize}

The above construction can easily be computed in polynomial time.
The next two lemmas show that it is indeed a desired reduction.

\begin{lemma}
If some set $S' \subseteq S$ with $|S'| \leq h$ contains at least
one element in each $S_j$, then there is a set $P \subseteq E'-E$
consisting of at most $p$ edges that holds 
 Properties N1 and N2.
\end{lemma}
\begin{proof}
For each $S_j$, let $s_{i_j}$ be an element in $S' \cap S_j$; by
the assumption of this lemma, these elements exist.  Next, let
$P_1=\{{b_1}\rightarrow{a_{i_1}},\ldots,
{b_r}\rightarrow{a_{i_r}}\}$ and
$P_2=\{{a_{i_1}}\rightarrow{b_0},\ldots,
{a_{i_r}}\rightarrow{b_0}\}$; by Rule 3, these two sets
exist. Now, let $P=P_1 \cup P_2 \cup \{{b_0}\rightarrow{a_0}\}$.
Note that $P \subseteq E'-E$. Since $P_1$ consists of $r$
edges and $P_2$ consists of at most $|S'|$ edges, $P$ has at most
$p$ edges.  $P$ holds   Property N1 because $E
\cup P$ consists of the edges in the traversable cycles
${b_0}\rightarrow{a_0},{a_0}\rightarrow{b_j},
{b_j}\rightarrow{a_{i_j}},{a_{i_j}}\rightarrow{b_0}$.  
Property N2 of $P$ follows from the fact that $P$ connects
$\{a_0,b_1,\ldots,b_r\}$, which forms the only connected
component of $\HH$ with more than one vertex.
\end{proof}

\begin{lemma}
If some set $P \subseteq E'-E$ consisting of at most $p$ edges
holds  Properties N1 and N2, then there
exists a set $S' \subseteq S$ with $|S'| \leq h$ that contains at
least one element in each $S_j$.
\end{lemma}
\begin{proof}
By  Property N1, $P$ must contain some edge
${b_j}\rightarrow{a_{i_j}}$ for each $j$ with $1 \leq j \leq
r$.  By Rule 3(4), $s_{i_j} \in S_j$.  Now let
$S'=\{s_{i_1},\ldots,s_{i_r}\}$.  To calculate the size of
$S'$, note that by  Property N1, $P$ must
also contain $b_0 \rightarrow a_0$ and at least one edge leaving
$a_{i_j}$ for each $j$.  Thus $|P| \geq |S'|+r+1$.  Then
$|S'| \leq h$ because $|P| \leq p=r+h+1$.
\end{proof}

The above lemma completes the proof of Theorem~\ref{thm_npc}.

\section{Optimal suppression in almost linear time}\label{sec_linear} 
Under the assumption that the total graph of $\TT$ is undirected,
this section considers the following optimization version of
Problem~\ref{problem_table_general}.

\begin{problem}[Optimal Suppression]\label{problem_table_undirected}
\rm
\begin{itemize}
\item Input: 
The suppressed graph $\HH=(A,B,E)$ of a table $\TT$ and a subset
$Q$ of $E$.
\item Output:
A set $P$ consisting of the smallest number of published cells in
$\TT$ such that $Q$ is totally protected in the table $\OT$
formed by $\TT$ with the cells in $P$ also suppressed.
\end{itemize}
\end{problem}

For all positive integers $n$ and $m$, let $\alpha$ denote the
best known function such that $m+n$ unions and finds of disjoint
subsets of an $n$-element set can be performed in
$O((m+n){\cdot}\alpha(n,m+n))$ time~\AlgTexts.

\begin{theorem}\label{thm_linear_time}
Problem~\ref{problem_table_undirected} can be solved in
$O((m+n){\cdot}\alpha(n,m+n))$ time, where $m$ is the number of
suppressed cells and $n$ is the total number of rows and columns
in $\TT$.
\end{theorem}

To prove Theorem~\ref{thm_linear_time},
Problem~\ref{problem_table_undirected} is first converted to the
next problem.

\begin{problem}\label{problem_graph_undirected}
\rm
\begin{itemize}
\item Input: 
An undirected bipartite graph $\HH=(A,B,E)$ and a subset $Q$ of
$E$.
\item Output: 
A forest $P$ formed by the smallest number of undirected edges
between $A$ and $B$ but not in $E$ such that the vertices of each
connected component of $(A,B,Q)$ are connected in $(A,B,(E-Q)
\cup P)$, i.e., contained in a connected component of $(A,B,(E-Q)
\cup P)$.
\end{itemize}
\end{problem}

\begin{lemma}\label{lem_reduce_undirected}
Problems~\ref{problem_table_undirected} and
{\ref{problem_graph_undirected}} can be reduced to each other in
linear time.
\end{lemma}
\begin{proof}
The proof uses arguments similar to those in the proof of
Lemma~\ref{lemma_equiv_general}. The strong connectivity
properties in Problem~\ref{problem_graph_general} and
Theorem~\ref{thm_test} can be ignored because this section
assumes that the total graph of $\TT$ is undirected. The forest
structure of $P$ follows from its minimality.
\end{proof}

Note that because $Q \subseteq E$, the vertices of each connected
component of $(A,B,Q)$ are connected in $(A,B,(E-Q) \cup P)$ if
and only if the vertices of each connected component of $\HH$ are
connected in $(A,B,(E-Q) \cup P)$.  Using this equivalence, the
next stage of the proof of Theorem~\ref{thm_linear_time} further
reduces Problem~\ref{problem_graph_undirected} to another graph
problem with the steps below:
\begin{list}{M\arabic{algcounterA}.}{
\usecounter{algcounterA}
\settowidth{\labelwidth}{M9.}
\settowidth{\leftmargin}{M9.}
\addtolength{\leftmargin}{\labelsep}}
\item
Compute the connected components $D_1,\cdots,D_r$ of $\HH$.
\item\label{a_step_maximal_E}
For each $D_i$, compute a maximal forest $K_i$ over the vertices
of $D_i$ using only the edges in $E-Q$.
\item\label{a_step_naive} 
For each $D_i$, extend $K_i$ to a maximal forest $L_i$ over the
vertices of $D_i$ using additional edges only from the complement
graph $D^c_i$ of $D_i$.
\item\label{step_contract_A}
Construct a graph $\hat{\HH}$ from $\HH$ by contracting each tree
in each $L_i$ into a single vertex.
\item
For each $D_i$, compute its contracted version $\hat{D}_i$ in
$\hat{\HH}$.
\item\label{last_step_A}
Divide the vertices of $\hat{\HH}$ into three sets, $V_A$, $V_B$,
$V_{AB}$, where a vertex in $V_A$ (respectively, $V_B$) consists
of a single vertex from $A$ (respectively, $B$), and a vertex in
$V_{AB}$ contains at least two vertices (thus with at least one
from each of $A$ and $B$).
\end{list}

A set of undirected edges between vertices in $V_A,V_B,V_{AB}$ is
called {\it semi-tripartite} if every edge in that set is between
two of the three sets or is between two vertices in $V_{AB}$.
Note that the set of edges in $\hat{\HH}$ is semi-tripartite.

\begin{problem}\label{problem_graph_2}
\rm
\begin{itemize}
\item Input: 
Three disjoint finite sets $V_A,V_B,V_{AB}$, and a partition
$\hat{D_1},\ldots,\hat{D_r}$ of $V_A \cup V_B \cup V_{AB}$.
\item Output:
A semi-tripartite set $\hat{P}$ consisting of the smallest number
of edges such that no edge in $\hat{P}$ connects two vertices in
the same $D_i$ and the vertices in each $D_i$ are connected in
the graph formed by is $\hat{P}$.
\end{itemize}
\end{problem}

\begin{lemma}\label{lemma_hat_P}
Problem~\ref{problem_graph_undirected} can be reduced to
Problem~\ref{problem_graph_2} in $O((m+n){\cdot}\alpha(n,m+n))$
time, where $m$ is the number of edges and $n$ is the number of
vertices in $\HH$.
\end{lemma}
\begin{proof}
The key idea is that an optimal $P$ for
Problem~\ref{problem_graph_undirected} can be obtained by
connecting the vertices of each $D_i$ first with edges in $E-Q$,
which can be used for free, next with edges in $D^c_i$, and then
with edges outside $D_i \cup D^c_i$.  Let $P'$ be a set of
$|\hat{P}|$ edges in the complement of $\HH$ that becomes
$\hat{P}$ after Step~M\ref{step_contract_A}.  Then,
$P'\cup(L_1-K_1)\cup\cdots\cup(L_r-K_r)$ is a desired output $P$
for Problem~\ref{problem_graph_undirected}, showing that Steps
M1--M\ref{last_step_A} can indeed reduce
Problem~\ref{problem_graph_undirected} to
Problem~\ref{problem_graph_2}.  Step M\ref{a_step_naive} is the
only step that requires more than linear time.  It is important
to avoid directly computing $D^c_i$ at Step M\ref{a_step_naive}.
Computing these complement graphs takes $\Theta(|A|{\cdot}|B|)$
time if some $D_i$ contains a constant fraction of the vertices
in $\HH$.  In such a case, if $\HH$ is sparse, then the time
spent on computing $D^c_i$ alone is far greater than the desired
complexity.  Instead of this naive approach,
Step~M\ref{a_step_naive} uses efficient techniques recently
developed for complement graph problems {\cite{KaoT94.isaac}} and
takes the desired $O((m+n){\cdot}\alpha(n,m+n))$ time.
\end{proof}

The last stage of the proof of Theorem~\ref{thm_linear_time} is
to give a linear-time algorithm for
Problem~\ref{problem_graph_2}.  A component $\hat{D_i}$ is {\it
good} if it has at least two vertices with at least one from
$V_{AB}$; it is {\it bad} if it has at least two vertices with
none from $V_{AB}$ (and thus with at least one from each of $V_A$
and $V_B$).  The goal is to use as few edges as possible to
connect the vertices in each of these components.  Let $w_g$ and
$w_b$ be the numbers of good and bad components, respectively.
There are three cases based on the value of $w_g$.

{\it Case} 1: $w_g = 0$.
If $w_b=0$, then let $\hat{P}=\emptyset$ because no $\hat{D_i}$
needs to be connected.
If $w_b > 0$ and $|V_{AB}| > 0$, then include in $\hat{P}$ an
edge between each vertex in the bad components and an arbitrary
vertex in $V_{AB}$.
If $w_b > 0$ and $|V_{AB}| = 0$, then there does not exist a
desired $\hat{P}$ and the given instance of
Problem~\ref{problem_graph_2} has no solution.

{\it Case} 2: $w_g = 1$.  Let $\hat{D}_j$ be the unique good component.

If $w_b > 0$, then find a bad component $\hat{D}_k$, and three
vertices $u \in V_{AB} \cap \hat{D}_j$, $v_1
\in V_A\cap\hat{D}_k$, $v_2 \in V_B\cap\hat{D}_k$.
Next, include in $\hat{P}$ an edge between $v_2$ and each vertex
in $(\hat{D}_j\cap(V_A \cup V_{AB}))-\{u\}$, an edge between
$v_1$ and each vertex in $\hat{D}_j \cap V_B$, and an edge
between $u$ and each vertex in the bad components.

If $w_b = 0$ and $V_{AB}-\hat{D}_j \neq \emptyset$, then include
in $\hat{P}$ an edge between every vertex in $\hat{D}_j$ and an
arbitrary vertex in $V_{AB}-\hat{D}_j$.

If $w_b = 0$ and $V_{AB}-\hat{D}_j=\emptyset$, then there are
sixteen subcases depending on whether $V_A \cap \hat{D}_j =
\emptyset$, $V_A - \hat{D}_j = \emptyset$, $V_B \cap \hat{D}_j =
\emptyset$, $V_B - \hat{D}_j = \emptyset$. If 
$V_A \cap \hat{D}_j\neq\emptyset$, $V_A-\hat{D}_j\not=\emptyset$,
$V_B\cap\hat{D}_j\not=\emptyset$, $V_B-\hat{D}_j\not=\emptyset$,
then include in $\hat{P}$ an edge between each vertex in $V_A
\cap \hat{D}_j$ and a vertex $v_2 \in V_B-\hat{D}_j$,
an edge between each vertex in $V_B \cap \hat{D}_j$ and a vertex
$v_1 \in V_A-\hat{D}_j$, and an edge between $v_1$ and each
vertex in $V_{AB} \cup \{v_2\}$. The other fifteen subcases are
handled similarly.

{\it Case} 3: $w_g \geq 2$.  Let $d$ be the total number of vertices in the
good and bad components.
Let $w'$ be the number of connected components in $\hat{P}$ that
contain the vertices of at least one good or bad $\hat{D}_i$; let
$d'$ be the number of vertices in these connected components of
$\hat{P}$ that are not in any good or bad $\hat{D}_i$.  By its
minimality, $\hat{P}$ forms a forest and $|\hat{P}| = d'+d-w'$.
The techniques for Cases 1 and 2 can be used to show that there
exists an optimal $\hat{P}$ with $d'=0$.  Thus, to minimize
$|\hat{P}|$ is to maximize $w'$.  Because two bad components
cannot be connected by edges between them alone, the strategy for
maximizing $w'$ is to pair a good component with a bad one,
whenever possible, and include in $\hat{P}$ edges between them to
connect their vertices into a tree.  After this step, if there
remain unconnected bad components but no unconnected good ones,
then add to $P$ an edge between each vertex in the remaining bad
components and an arbitrary vertex in the intersection of
$V_{AB}$ and a good component.  On the other hand, if there
remain good components but no bad ones, then pair up these good
components similarly.  After this step, if there remains a good
component, then add to $\hat{P}$ an edge between each vertex in
this last good component and an arbitrary vertex in the
intersection of $V_{AB}$ and another good component.  (As a
result, if $w_g \leq w_b$, then $|\hat{P}| = d-w_g$; otherwise,
$|\hat{P}|=d-\lfloor{{w_g+w_b}\over{2}}\rfloor$.)

The above discussion yields a linear-time algorithm for
Problem~\ref{problem_graph_2} in a straightforward manner. This
finishes the proof of Theorem~\ref{thm_linear_time}.

\section{Discussions}\label{sec_extension}
Lemma~\ref{lem_reduce_undirected} has several significant
implications.  Since $P$ is a forest, it has at most $n-1$ edges.
Thus, for a table with an undirected total graph, no more than
$n-1$ additional cells need to be suppressed to achieve total
protection. This is a small number compared to the size of the
table, which may grow quadratically in $n$. Moreover, when $\HH$
is connected and $E=Q$, $(A,B,P)$ is a spanning tree.  In this
case, many well-studied tree-related computational concepts and
tools, such as minimum-cost spanning trees, can be applied to
consider other optimal suppression problems for total protection.

\section*{Acknowledgements}
The author is deeply grateful to Dan Gusfield for his constant
encouragement and help.  The author wishes to thank an anonymous referee
for very helpful and thorough comments. The referee has also pointed out
that some very interesting materials related to Theorems~\ref{thm_mli} and
\ref{thm_uk_bk_lem_unbounded_embedding} have been developed in the context
of protecting sums of suppressed cells
\cite{Malvestuto93, MM90, MMR91}.

%\bibliographystyle{siam}
%\bibliography{all}

\end{document}